\documentclass[10pt,twocolumn,letterpaper]{article} 

\usepackage[pagenumbers]{cvpr} 

\usepackage{graphicx}
\usepackage{amsmath}
\usepackage{amssymb}
\usepackage{booktabs}
\usepackage{bm}
\usepackage{makecell}
\usepackage{booktabs}
\usepackage{multirow}
\usepackage{color}
\usepackage{pifont}

\usepackage[pagebackref,breaklinks,colorlinks]{hyperref}

\usepackage[capitalize]{cleveref}
\crefname{section}{Sec.}{Secs.}
\Crefname{section}{Section}{Sections}
\Crefname{table}{Table}{Tables}
\crefname{table}{Tab.}{Tabs.}

\def\cvprPaperID{2802} 
\def\confName{CVPR}
\def\confYear{2022}

\begin{document}
	
	\title{Details or Artifacts: A Locally Discriminative Learning Approach to \\Realistic Image Super-Resolution}
	
	\author{\textbf{Jie Liang}$^{1}$\footnotemark[1],\;  \textbf{Hui Zeng}$^{2}$\footnotemark[1]\; and \textbf{Lei Zhang}$^{1}$\footnotemark[2]\\
		$^1$The HongKong Polytechnic University,\;  $^2$OPPO Research\\
		\textit{\{liang27jie,\, cshzeng\}@gmail.com}; \textit{cslzhang@comp.polyu.edu.hk}\\
	}

	\maketitle
	
	\renewcommand{\thefootnote}{\fnsymbol{footnote}}
	\footnotetext[1]{Equal contribution.}
	\footnotetext[2]{This work is supported by the Hong Kong RGC RIF grant (R5001-18).}
	
	\begin{abstract}
		Single image super-resolution (SISR) with generative adversarial networks (GAN) has recently attracted increasing attention due to its potentials to generate rich details. However, the training of GAN is unstable, and it often introduces many perceptually unpleasant artifacts along with the generated details. In this paper, we demonstrate that it is possible to train a GAN-based SISR model which can stably generate perceptually realistic details while inhibiting visual artifacts. Based on the observation that the local statistics (e.g., residual variance) of artifact areas are often different from the areas of perceptually friendly details, we develop a framework to discriminate between GAN-generated artifacts and realistic details, and consequently generate an artifact map to regularize and stabilize the model training process. Our proposed locally discriminative learning (LDL) method is simple yet effective, which can be easily plugged in off-the-shelf SISR methods and boost their performance. Experiments demonstrate that LDL outperforms the state-of-the-art GAN based SISR methods, achieving not only higher reconstruction accuracy but also superior perceptual quality on both synthetic and real-world datasets. Codes and models are available at  \href{https://github.com/csjliang/LDL}{https://github.com/csjliang/LDL}.
	\end{abstract}
	
	\section{Introduction}
	
	\begin{figure}[t]
		\centering
		\includegraphics[width=0.47\textwidth]{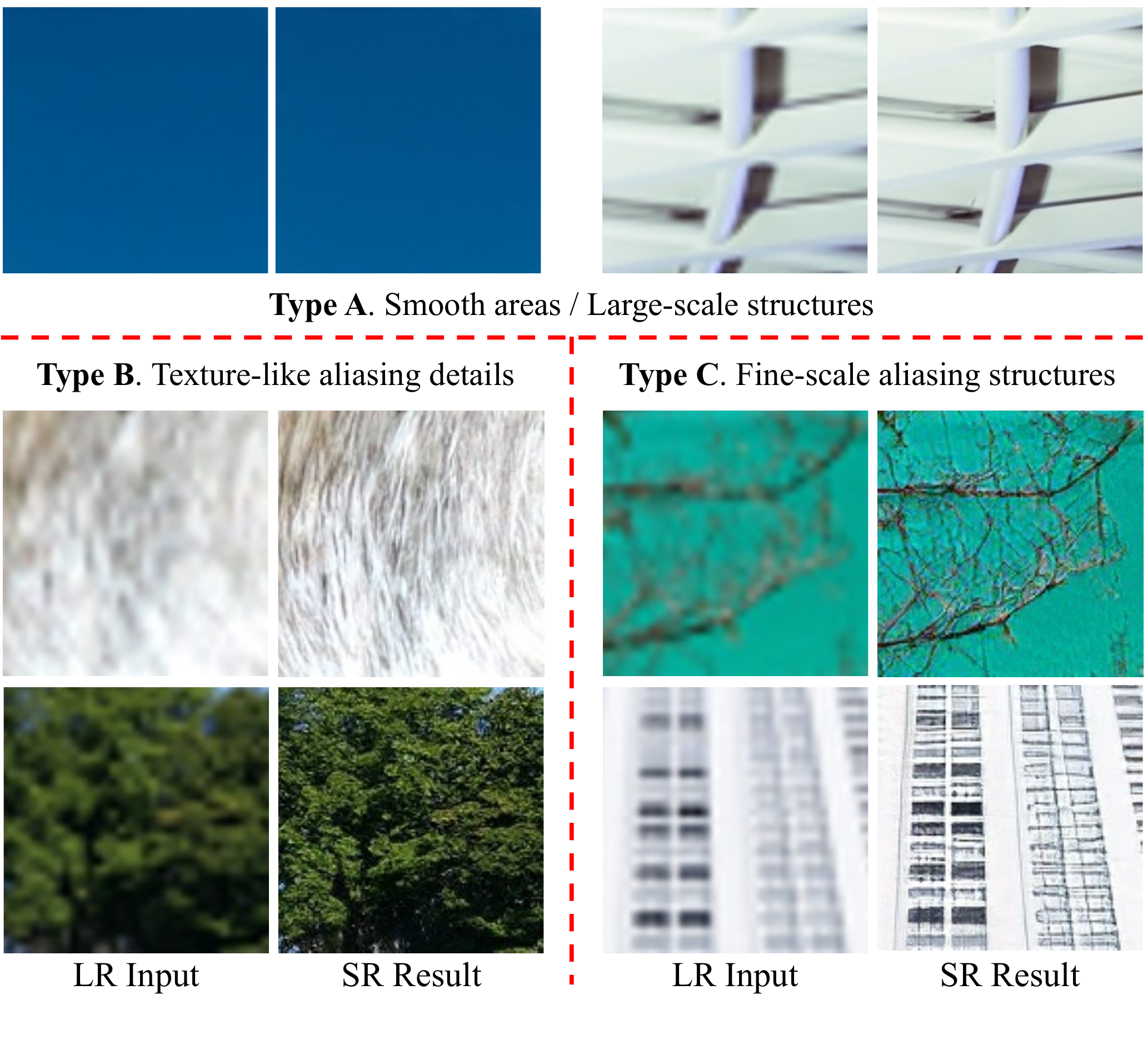}
		\vspace{-1em}
		\captionof{figure}{\label{introduction}Three representative types of SISR regions generated by ESRGAN~\cite{wang2018esrgan}. For each example, the left is an LR patch and the right is its GAN-SR result. Type A patches represent regions that are easy to super-resolve, \eg, smooth and large-scale structural areas, where the main structures are preserved in the LR input. In contrast, patches of type B and type C are with fine-scale details, which are hard to be faithfully restored due to the signal aliasing in the LR inputs. The results of texture-like type B patches are perceptually realistic despite the pixel-wise differences to the ground-truth, since the patterns are naturally irregular with weak priors for observers. However, the results of type C patches exhibit perceptually unpleasant visual artifacts since the overshoot pixels and distorted structures are sensitive to human perception.}\vspace{-1em}
	\end{figure}
	
	Single image super-resolution (SISR)~\cite{dong2014learning, johnson2016perceptual, ledig2017photo, sajjadi2017enhancenet, soh2019natural, kim2016deeply, sun2010gradient, zhang2021designing, wang2021realesrgan, shi2016real, wang2018esrgan, yan2015single, zhang2018image, zhang2018residual, jo2021tackling}, which aims to reconstruct a high-resolution (HR) image from its low-resolution (LR) observation, is one hot yet challenging research topic in low-level computer vision. It has become prevalent to train deep neural networks (DNNs) for SISR, while many DNN-based SISR models~\cite{dong2014learning, zhang2018residual, anwar2019drln, niu2020single, wang2021learning} are trained with pixel-wise $\ell_1$ and $\ell_2$ losses, and/or local window based metrics (such as SSIM~\cite{wang2004image}). It is well-known that though high PSNR and SSIM indices might be induced by these losses, they can hardly produce rich image details~\cite{blau2018perception, ledig2017photo}.
	
	With the rapid development of generative adversarial networks (GAN)~\cite{goodfellow2014generative, jolicoeur2018relativistic}, GAN-based SISR (GAN-SR for short) has recently attracted significant attention for its potentials to recover sharp images with rich details~\cite{ledig2017photo, sajjadi2017enhancenet, wang2018esrgan, soh2019natural, zhang2020deep}. Though great progresses have been achieved, adversarial training is unstable and often introduces unpleasant visual artifacts~\cite{ledig2017photo, zhang2020deep}. As users are mostly expecting \textit{rich and realistic details} in SISR results~\cite{prashnani2018pieapp, ding2020image, jinjin2020pipal}, how to inhibit the visual artifacts of GAN-SR without affecting the realistic details becomes a key issue. Unfortunately, details and artifacts are often entangled in high-frequency components of images. As a result, optimizing one of them often harms the other under existing frameworks~\cite{blau2018perception, ledig2017photo, wang2018esrgan, ma2020structure}.

	In order to address the above mentioned challenges, we investigate in-depth the GAN-SR methods and categorize their results into three typical types of regions, as illustrated in Figure~\ref{introduction}. Specifically, type A patches (\eg, flat sky, long edges) are easy to reconstruct since they are smooth or contain only large-scale structures. In contrast, it is difficult to produce high-fidelity SISR results for patches of type B and type C because they have much fine-scale details and suffer from signal aliasing in the degradation process, where most high-frequency components are lost. Fortunately, for texture-like type B patches (\eg, animal fur, tree leaves in distance), the pixels are randomly distributed so that the differences between SR results and ground truth are insensitive to human perception. Therefore, rich details generated by GAN-SR methods can lead to better perceptual quality in these regions. However, patches of type C (\eg, thin twigs, dense windows in the building) contain many fine-scale regular structures or sharp transitions among adjacent pixels. The distorted structures and overshoot pixels generated by GAN-SR methods can be easily perceived by observers as unpleasant artifacts.
	
	Based on the above analysis, we can see that to get perceptually realistic SISR results, the visual artifacts in type C regions should be inhibited, while the realistic details generated in type A and type B regions should be preserved. To achieve this goal, we analyze the local statistic of the three types of GAN-SR regions, and find that the local variance of residuals between SISR results and ground truth HR images can serve as an effective feature to distinguish unpleasant artifacts from realistic details. Accordingly, we construct a pixel-wise map indicating the probability of each pixel being artifacts based on the local and patch-level residual variances. We further refine the discrimination map via a model ensemble strategy to encourage stable and accurate optimization direction toward high-fidelity reconstruction. Based on the refined map, we design a Locally Discriminative Learning (LDL) framework to penalize the artifacts without affecting realistic details.
	
	To sum up, in this paper we first analyze the GAN-SR results and the instability of model training. We then propose to explicitly discriminate visual artifacts from realistic details, and design an LDL framework to regularize the adversarial training. Our method is simple yet effective, and it can be easily plugged into off-the-shelf GAN-SR methods. It provides a novel way to suppress the artifacts in GAN-SR while generating rich realistic details. We conduct extensive experiments on synthetic and real-world SISR tasks, and LDL demonstrates clear improvements against the state-of-the-arts both quantitatively and qualitatively.

	\section{Related work}
	
	Since the pioneer work of SRCNN~\cite{dong2014learning}, which firstly introduces a three-layer convolutional neural network (CNN) for SISR, a number of CNN based SISR models have been proposed, which can be roughly divided into signal fidelity-oriented ones~\cite{zhang2018residual, anwar2019drln, niu2020single, wang2021learning} and perceptual quality-oriented ones~\cite{johnson2016perceptual, ledig2017photo, sajjadi2017enhancenet, soh2019natural, wang2018esrgan, liang2021hierarchical}, depending on the losses and training strategies employed by them.    
	
	\textbf{Signal fidelity-oriented SISR methods}. SISR methods in this category adopt the pixel-wise distance measures (such as $\ell_2$ and $\ell_1$ losses) and local structural similarity measures (such as SSIM~\cite{wang2004image}) to optimize the signal fidelity between the SISR outputs and the HR ground-truth. Since SRCNN~\cite{dong2014learning}, researchers have made remarkable progresses by stacking more convolution layers~\cite{kim2016accurate, kim2016deeply} and designing more complex building blocks~\cite{lim2017enhanced, tai2017memnet} and connections~\cite{ledig2017photo, zhang2018residual, tong2017image}. For instance, benefited from the very deep network,  effective residual connections, and channel attentions, RCAN~\cite{zhang2018image} achieves superior performance on reconstruction accuracy (\eg, PSNR). However, due to the ill-posedness of the SISR problem, optimizing the pixel-wise losses tends to find a blurry result that is the average of many possible solutions~\cite{sajjadi2017enhancenet, soh2019natural, blau2018perception}. The SSIM loss can preserve better the image local structures but it is hard to reproduce fine details.
	
	\textbf{Perceptual quality-oriented SISR methods}.	To improve the perceptual quality of SISR images, Johnson~\etal~\cite{johnson2016perceptual} proposed a perceptual loss by calculating the distance between HR and SISR results in the VGG feature space. To tackle the difficulties of signal fidelity-oriented methods in reproducing image details, most recent works have resorted to using the GAN techniques~\cite{goodfellow2014generative} for their capability to generate desired images by discriminating between image distributions~\cite{wang2018recovering, rad2019srobb, fuoli2021fourier}. For example, Ledig~\etal~\cite{ledig2017photo} proposed SRGAN with adversarial training on top of the SRResNet generator. To improve the visual quality, Wang~\etal~\cite{wang2018esrgan} proposed the ESRGAN by introducing the Residual-in-Residual Dense Block (RRDB) along with other improvements on adversarial training and perceptual loss. RRDB has been employed as a standard backbone in many state-of-the-art GAN-SR methods~\cite{wang2021realesrgan, zhang2021designing, ma2020structure}.
	
	\begin{figure}[t]
		\centering
		\includegraphics[width=0.4\textwidth]{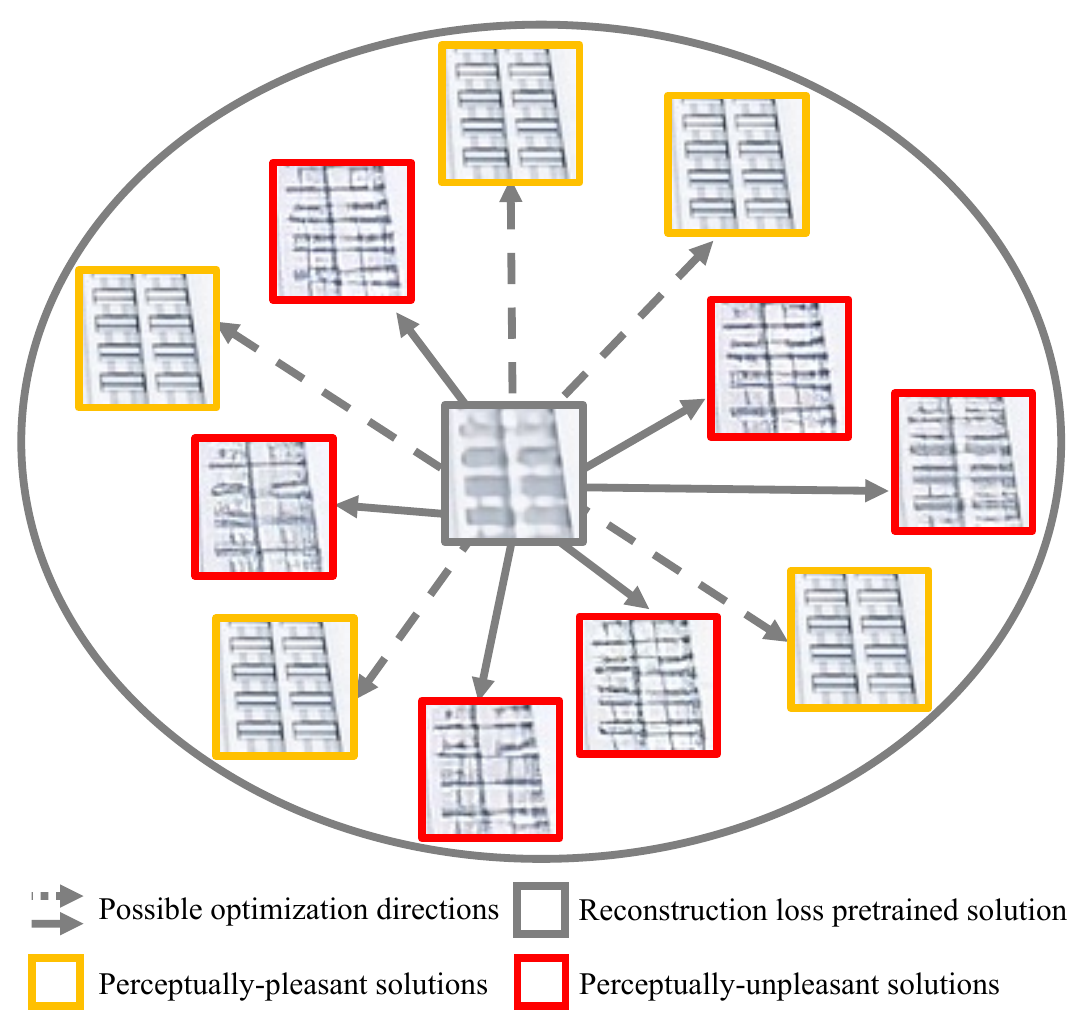}
		\vspace{-1em}
		\captionof{figure}{\label{directions}An illustration of possible optimization directions of GAN-SR models. The patch in the center is obtained by a pre-trained SISR model using $\ell_1$-loss, while the patches in red and yellow boxes are possible GAN-SR results by adversarial losses.}\vspace{-1em}
	\end{figure}

	Zhang~\etal~\cite{zhang2020deep} proposed a trainable unfolding network, termed USRGAN, which integrates the merits of traditional model-based methods and CNN-based ones. Ma~\etal~\cite{ma2020structure} introduced a gradient guidance via an additional branch in the network. By alleviating the structural distortion and inconsistency problem, the proposed SPSR method achieves leading performance among GAN-SR methods on synthetic data. Nonetheless, one key issue of all existing GAN-SR works lies in that they will produce many unpleasant visual artifacts due to the instability of adversarial training. 
	
	\textbf{Remarks}. As indicated in~\cite{blau2018perception}, both signal fidelity- and perceptual quality-oriented SISR methods fall in a perception-distortion trade-off; that is, improving either the perceptual quality or signal fidelity will affect the other under the existing training strategies. Empirical experiences also tell us that inhibiting the artifacts can limit the generation of details. In this paper, we propose to regularize the adversarial training by explicitly discriminating the artifacts from realistic details, which effectively addresses the dilemma. Recent researches, \eg, BSRGAN~\cite{zhang2021designing} and RealESRGAN~\cite{wang2021realesrgan}, have also recognized the significance of the real-world image SR task. As a plug-and-play module, our method can also be easily extended to such challenging task. The experimental results demonstrated its high generalization performance in generating realistic details while inhibiting artifacts.

	\section{Methodology}
	
	\subsection{GAN-SR induced visual artifacts}\label{investigations}
	
	\begin{figure}[t]
		\centering
		\includegraphics[width=0.42\textwidth]{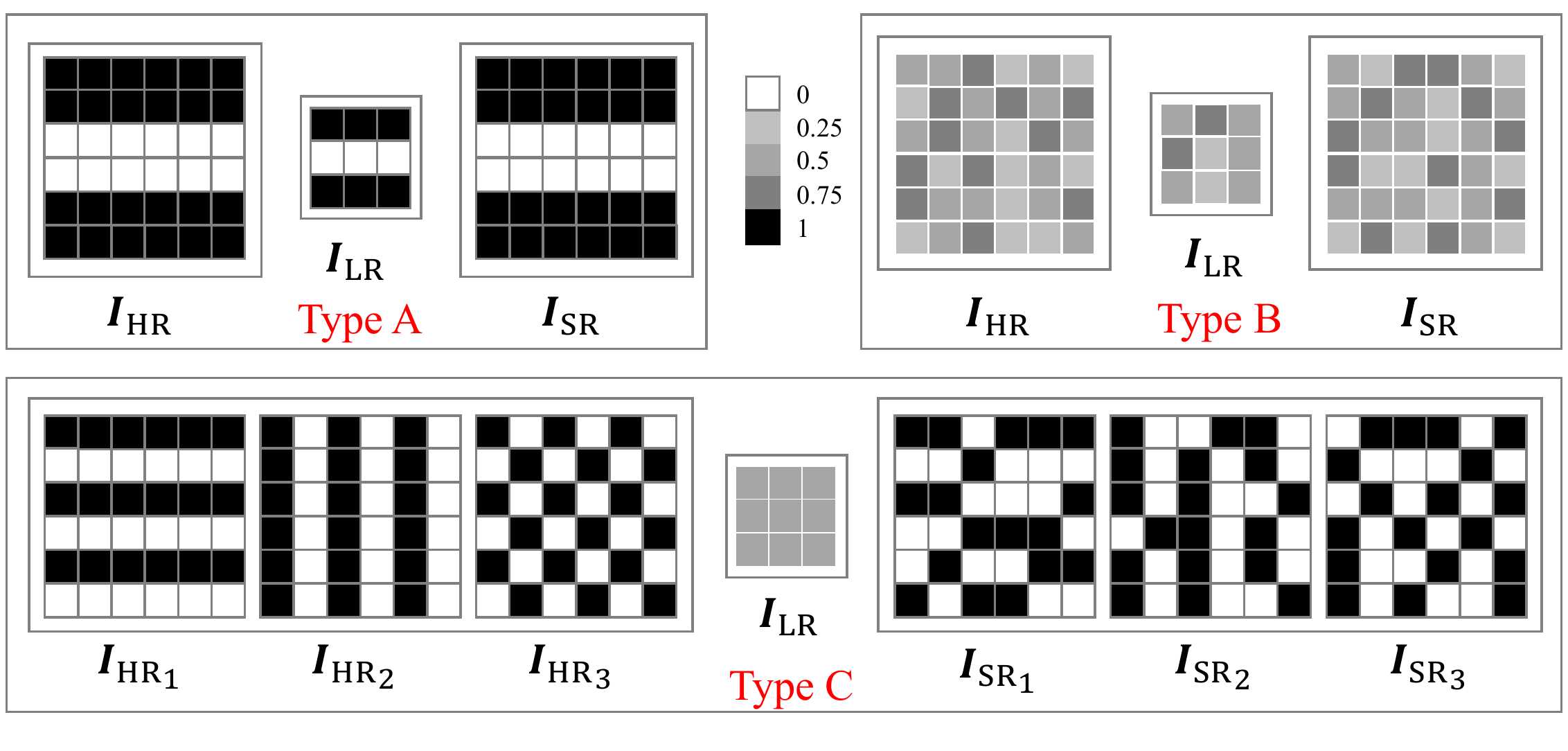}
		\vspace{-1em}
		\captionof{figure}{\label{toy} Toy examples of the GAN-SR results on three types of regions. The LR patches are obtained by applying $2\times2$ average pooling with stride $2$ on the HR patches. The large-scale structure in type A patch can be well reproduced with good fidelity and perceptual quality. Though the pixels in texture-like type B patch are not faithfully reconstructed, the perceptual quality of the reconstructed patch is not bad due to the random distribution of pixels in HR patch. However, for those type C patches, visually unpleasant artifacts are perceived in the GAN-SR results since the fine-scale yet regular structures are destroyed.}\vspace{-1em}
	\end{figure}
	
	Most of the existing GAN-SR methods~\cite{ledig2017photo, wang2018esrgan} are trained using a weighted combination of three losses:	
	\begin{equation}
		\label{GAN_loss}
		\mathcal{L}_{\text{GAN}} = \lambda_1\mathcal{L}_{\text{recons}} + \lambda_2\mathcal{L}_{\text{percep}} + \lambda_3\mathcal{L}_{\text{adv}},
	\end{equation}
	where $\mathcal{L}_{\text{recons}}$ indicates the pixel-wise reconstruction loss such as $\ell_1$ and $\ell_2$ distances, $\mathcal{L}_{\text{percep}}$ is the perceptual loss~\cite{johnson2016perceptual, ledig2017photo} measuring the feature distance in VGG feature space and $\mathcal{L}_{\text{adv}}$ denotes the adversarial loss~\cite{goodfellow2014generative, wang2018esrgan}. $\lambda_1, \lambda_2$ and $\lambda_3$ are balancing parameters, which are usually set to $0.01, 1, 0.005$, respectively, as in ESRGAN~\cite{wang2018esrgan}.
	
	According to the pioneer work of SRGAN~\cite{ledig2017photo}, using only the $\mathcal{L}_{\text{recons}}$ loss will result in a blurred average of all possible HR images, while the $\mathcal{L}_{\text{adv}}$ loss can push the SISR solution away from the blurred average, generating more details. Unfortunately, GAN-SR models also generate many perceptually-unpleasant artifacts in addition to the details. An intuitive illustration is shown in Figure~\ref{directions}. Since SISR is an ill-posed task, one LR input corresponds to many possible HR counterparts scattering in the high-dimensional image space. Starting from the blurry solution (the center patch in Figure~\ref{directions}) generated by an SISR model pre-trained using only the $\mathcal{L}_{\text{recons}}$ loss, the $\mathcal{L}_{\text{GAN}}$ loss can update it along many possible directions, some yielding perceptually pleasant results (in yellow boxes) and some producing unpleasant ones (in red boxes). This leads to an unstable optimization process that may generate artifacts along with details.

	The above situation can vary among different image regions, as discussed in Figure~\ref{introduction}. To better understand how GAN-SR generates visual artifacts in different areas of an image, in Figure~\ref{toy} we show toy examples of the three types of patches. We see that for type A patch, the large-scale structure is preserved in its LR version and the HR patch can be easily reproduced with good fidelity and perceptual quality. For the texture-like type B patch, though it is not pixel-wise faithfully reconstructed, the perceptual quality of the GAN-SR output is not bad. This is mainly because the pixels in texture-like patches are often randomly distributed in a relatively small range so that human eyes are hard to perceive the pixel-wise difference. In contrast, type C patches have regular and sharp transitions, while the local patterns are lost in the LR patch after degradation. The largely varied and even contradictory HR targets lead to unstable adversarial training, and the irregular and unnatural patterns in the GAN-SR results can be easily perceived by observers as artifacts.

\begin{figure}[t]
		\centering
		\includegraphics[width=0.42\textwidth]{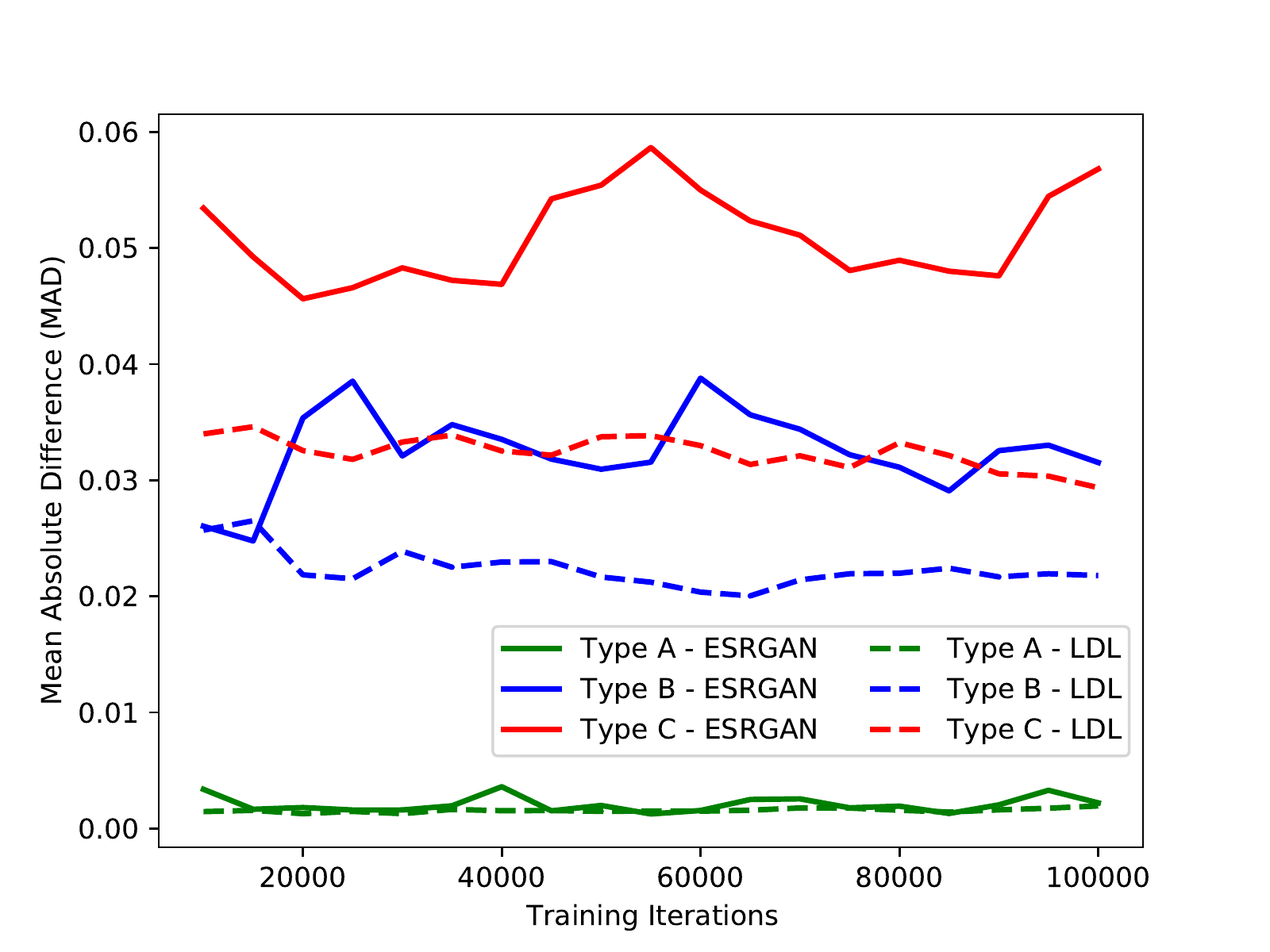}
		\vspace{-1em}
		\captionof{figure}{\label{curve} The stability on the training of different patches by ESRGAN~\cite{wang2018esrgan} and our LDL. The patches of flat sky (type A), animal fur (type B) and thin twigs (type C) in Figure~\ref{introduction} are used here. The mean absolute differences (MAD) of intermediate GAN-SR results between iterations \textit{k} and \textit{k}+5000 are plotted.}\vspace{-1em}
	\end{figure}

	In Figure~\ref{curve}, we further investigate the training stability of GAN-SR methods on different patches, including the flat sky (type A), animal fur (type B) and thin twigs (type C) in Figure~\ref{introduction}. We calculate the mean absolute difference (MAD) of the intermediate GAN-SR outputs at two different iterations, \ie, MAD=$|\bm{I}_{\text{SR}}^{(k)} - \bm{I}_{\text{SR}}^{(k+p)}|$, where $\bm{I}_{\text{SR}}^{(k)}$ is the GAN-SR result at iteration $k$, and we set $p$ to $5000$. The curves of MAD vs. $k$ for ESRGAN~\cite{wang2018esrgan} are plotted as solid lines. As can be seen, the training process of type A patch is stable (small value and variation of MAD). Type B shows larger variation, indicating higher uncertainty during optimization. Type C has the largest variation and instability, implying that many possible GAN-SR solutions of type C are available in a large space, as illustrated in Figure~\ref{directions}.
	
	\subsection{Discriminating artifacts from realistic details}
	
\begin{figure*}[t]
		\centering
		\includegraphics[width=0.955\textwidth]{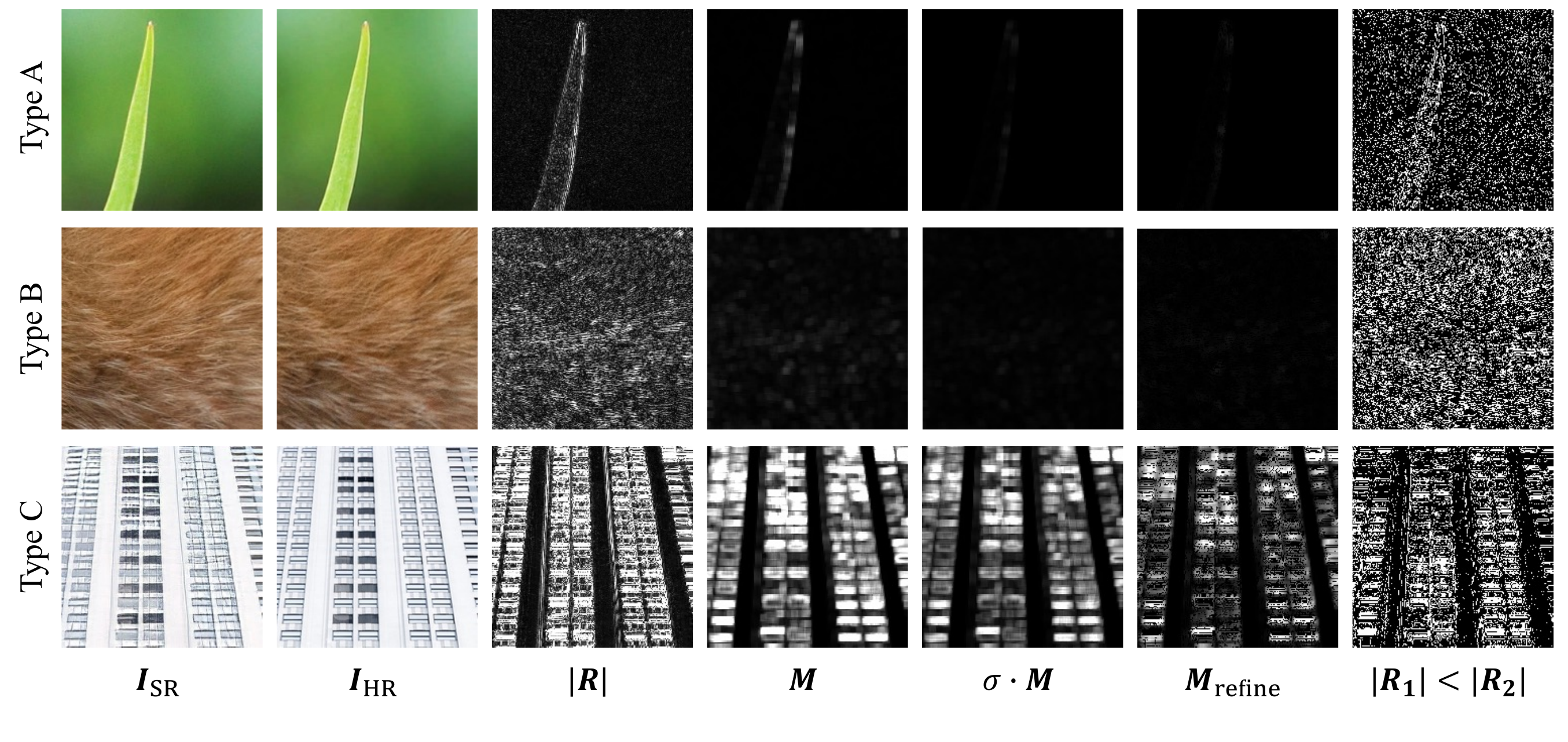}
		\vspace{-2em}
		\captionof{figure}{\label{var_valid}Visualization on the generation process of artifact map. $\bm{I}_{\text{SR}}, \bm{I}_{\text{HR}}, |\bm{R}|, \bm{M}, \sigma$ and $\bm{M}_{\text{refine}}$ indicate the SISR output of a GAN-SR method, the ground truth patch, the absolute value of the residual between $\bm{I}_{\text{SR}}$ and $\bm{I}_{\text{HR}}$, the primary map calculated by Eq.~\eqref{var_local}, the scaling factor computed by Eq.~\eqref{var_global}, and the refined map by Eq.~\eqref{refine}, respectively. In the $5^{th}$ column, the $\sigma$ values for type A, B and C patches are $0.25, 0.39, 0.67$, respectively. The last column shows the locations where $|\bm{R}_1|<|\bm{R}_2|$ with white pixels.}\vspace{-1em}
	\end{figure*}

	According to the investigations in Section~\ref{investigations}, we should inhibit the generation of artifacts in type C patches while preserving the realistic details in type A and B patches. To achieve this challenging goal, we carefully design a pixel-wise map to discriminate artifacts from realistic details, as well as a learning strategy to stabilize the training of GAN-SR models. The whole procedure of map generation is illustrated in Figure~\ref{var_valid} using three patches.
	
	\textbf{Discrimination of artifacts.} Suppose that the resolution of a full-color SISR image $\bm{I}_{\text{SR}}$ is ${H\times W\times 3}$, our goal is to find a pixel-wise map $\bm{M}\in\mathbb{R}^{H\times W\times 1}$, where $\bm{M}({i,j})\in [0,1]$ indicates the probability of $\bm{I}_{\text{SR}}(i,j)$ being an artifact pixel. Considering that both the artifacts and details belong to high-frequency image components, we first calculate the residual between ground truth image $\bm{I}_{\text{HR}}$ and SISR result $\bm{I}_{\text{SR}}$ to extract high-frequency components:
	\vspace{-0.2em}
	\begin{equation}
		\label{residual}
		\bm{R} = \bm{I}_{\text{HR}} - \bm{I}_{\text{SR}}.\vspace{-0.2em}
	\end{equation}
	
	As shown in the $3^{rd}$ column of Figure~\ref{var_valid}, most pixels in the smooth type A patch have very small residuals. Both type B and type C patches have large residuals, while the distribution of residuals in patch B is much more random. Based on the observation that artifacts usually consist of overshoot pixel values, we propose to calculate the local variance of the residual map $\bm{R}$ as the primary map to indicate artifact pixels:
	\vspace{-0.3em}
	\begin{equation}
		\label{var_local}
		\bm{M}(i,j) = var(\bm{R}(\text{\scriptsize $i$}-\text{\scriptsize $\frac{n-1}{2}$}:\text{\scriptsize $i$}+\text{\scriptsize $\frac{n-1}{2}$}, \text{\scriptsize $j$}-\text{\scriptsize $\frac{n-1}{2}$}:\text{\scriptsize $j$}+\text{\scriptsize $\frac{n-1}{2}$}),
	\end{equation}
	where $var$ represents the variance operator and $n$ denotes the local window size. We empirically set $n=7$.
	
	As shown in the $4^{th}$ column of Figure~\ref{var_valid}, the primary map $\bm{M}$ can effectively detect the artifact pixels in patch C. However, since the local variance is calculated with a very small receptive field, it is unstable to discriminate artifacts from edges and textures. Some pixels in patches A and B will also have large response, causing wrong punishment on the generation of realistic details. To address this issue, we further calculate a stable patch-level variance $\sigma$ from the whole residual map $\bm{R}$ as follows:
	\vspace{-0.2em}
	\begin{equation}
		\label{var_global}
		\sigma = (var(\bm{R}))^{\frac{1}{a}},\vspace{-0.2em}
	\end{equation}
	where $(\cdot)^{\frac{1}{a}}$ scales the global variance $var(\bm{R})$ to an appropriate scale. We fix $a$ to $5$ throughout our experiments. In general, type A patches have smaller $\sigma$ values than type B and type C patches, while type C patches have the largest $\sigma$ values. By using $\sigma$ to scale the primary map $\bm{M}$ as $\sigma\cdot\bm{M}$, a more reliable artifact map can be obtained. As shown in the $5^{th}$ column of Figure~\ref{var_valid}, the over-punishment issue on patches A and B is mostly addressed, while the artifacts in patch C are still identified.
	
	\textbf{Stabilization and refinement.} Although the map $\sigma\cdot\bm{M}$ can discriminate the artifacts in different types of patches, it may still over-penalize the realistic details in patch C, and slightly penalize the generation of high-fidelity details in patches A and B, especially at the early training stages. To alleviate this problem, we further stabilize the training process and refine the artifact map. 
	
	Specifically, denote by $\Psi$ the GAN-SR model optimized via gradient decent on-the-fly, we use the exponential moving average (EMA) technique to temporally ensemble a more stable model $\Psi_{\text{EMA}}$ from $\Psi$ as:
	\vspace{-0.2em}
	\begin{equation}
		\label{ema}
		\Psi_{\text{EMA}}^{(k)} = \alpha \cdot \Psi_{\text{EMA}}^{(k-1)} + (1 - \alpha) \cdot \Psi^{(k)}, \vspace{-0.2em}
	\end{equation}
	\noindent where $\alpha$ is the weighting parameter. Compared to $\Psi$, $\Psi_{\text{EMA}}$ is more reliable to alleviate the generation of random artifacts. As in prior arts of EMA~\cite{karras2019style, karras2020analyzing}, we set $\alpha=0.999$. 
	
	With $\Psi_{\text{EMA}}$, we can further refine the artifact map $\sigma\cdot\bm{M}$ to alleviate penalty on generation of realistic details during optimization. Denote by $\bm{I}_{\text{SR}_1} = \Psi(\bm{I}_{\text{LR}})$ and $\bm{I}_{\text{SR}_2} = \Psi_{\text{EMA}}(\bm{I}_{\text{LR}})$ the outputs of two GAN-SR models. Usually, the output of the ensemble model, \ie, $\bm{I}_{\text{SR}_2}$, has few artifacts, while $\bm{I}_{\text{SR}_1}$ may contain more details and artifacts simultaneously. We then calculate two residuals map $\bm{R}_1 = \bm{I}_{\text{HR}} - \bm{I}_{\text{SR}_1}$ and $\bm{R}_2 = \bm{I}_{\text{HR}} - \bm{I}_{\text{SR}_2}$, and refine the artifact map $\sigma\cdot\bm{M}$ by:
	\vspace{-0.2em}
	\begin{equation}
		\label{refine}
		\bm{M}_{\text{refine}}(i,j) = 
		\begin{cases}
			\text{\footnotesize $0$}, &\text{\footnotesize if $|\bm{R}_{1}(i,j)| < |\bm{R}_{2}(i,j)|$};\\
			\text{\footnotesize $\sigma\cdot\bm{M}(i,j)$}, &\text{\footnotesize if $|\bm{R}_{1}(i,j)| \geq |\bm{R}_{2}(i,j)|$}.
		\end{cases}\vspace{-0.2em}
	\end{equation}
	That is, the refined map $\bm{M}_{\text{refine}}$ will only penalize the pixels where $|\bm{R}_{1}(i,j)| \geq |\bm{R}_{2}(i,j)|$. At locations where the residuals of $\bm{I}_{\text{SR}_1}$ are smaller than $\bm{I}_{\text{SR}_2}$, the model $\Psi$ is updated towards the correct direction and should not be penalized. The refined map $\bm{M}_{\text{refine}}$ and the location map $|\bm{R}_{1}| < |\bm{R}_{2}|$ are shown in the last two columns of Figure~\ref{var_valid}. We see that the locations of fine textures and desirable edges are removed from the refined artifact map so that the penalty can be imposed more precisely on the artifact pixels.

	\subsection{Loss and learning strategy}
	
	Given the refined artifact map $\bm{M}_{\text{refine}}$, we propose an artifact discrimination loss $\mathcal{L}_{\text{artif}}$ as follows:
	\vspace{-0.2em}
	\begin{equation}
		\mathcal{L}_{\text{artif}} = \lVert \bm{M}_{\text{refine}}\cdot(\bm{I}_{\text{HR}} - \bm{I}_{\text{SR}_1})\rVert_1.\vspace{-0.2em}
	\end{equation}
	The loss $\mathcal{L}_{\text{artif}}$ can be easily introduced to the existing GAN-SR models and the final loss function is:
	\vspace{-0.2em}
	\begin{equation}
		\mathcal{L}_{\text{LDL}} = \mathcal{L}_{\text{GAN}} + \beta\mathcal{L}_{\text{artif}},\vspace{-0.2em}
	\end{equation}
	where $\mathcal{L}_{\text{GAN}}$ is defined in Eq.~\eqref{GAN_loss} and $\beta$ is a weighting parameter. We simply fix $\beta=1$ in all our experiments.
	
	\begin{figure}[t]
		\centering
		\includegraphics[width=0.35\textwidth]{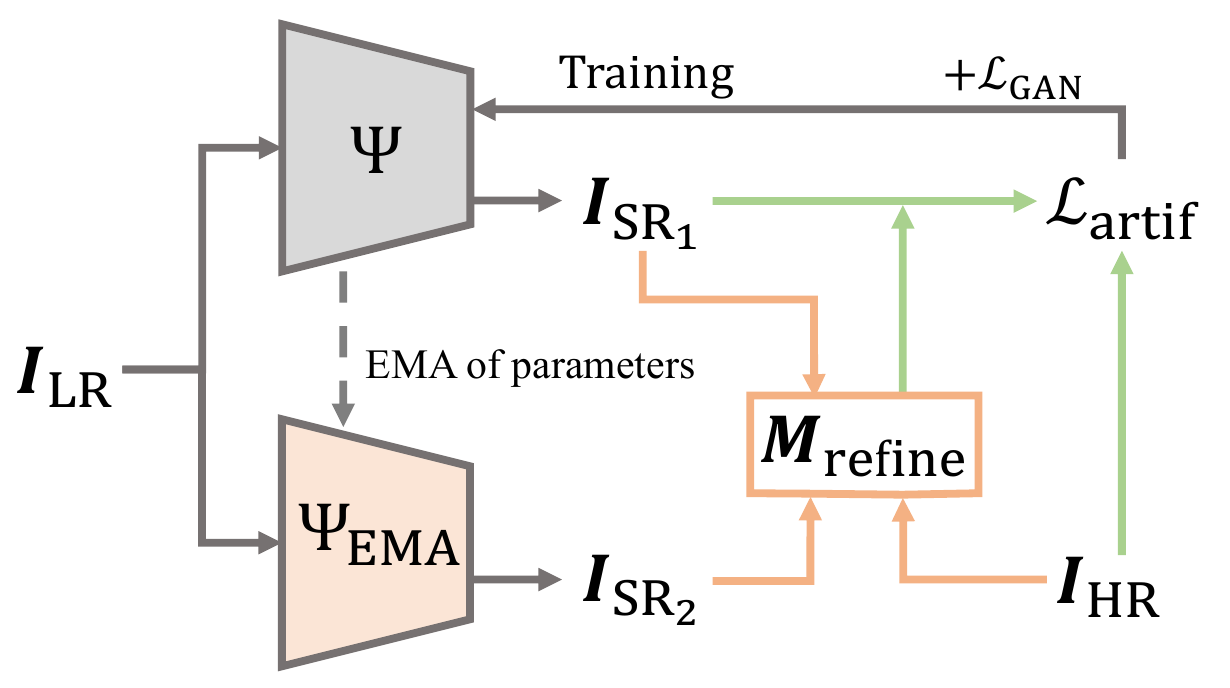}
		\vspace{-0.8em}
		\captionof{figure}{\label{pipeline}Overall learning pipeline of the proposed LDL method.}\vspace{-1em}
	\end{figure}
	
	The pipeline of the proposed locally discriminative learning (LDL) method is shown in Figure~\ref{pipeline}. The input $\bm{I}_{\text{LR}}$ is fed into two models, \ie, $\Psi$ and $\Psi_{\text{EMA}}$, to output $\bm{I}_{\text{SR}_1}$ and $\bm{I}_{\text{SR}_2}$, respectively. The artifact map $\bm{M}_{\text{refine}}$ is then constructed using the ground-truth image $\bm{I}_{\text{HR}}$, as well as $\bm{I}_{\text{SR}_1}$ and $\bm{I}_{\text{SR}_2}$. After that, the loss $\mathcal{L}_{\text{artif}}$ is calculated based on $\bm{I}_{\text{HR}}$, $\bm{I}_{\text{SR}_1}$ and $\bm{M}_{\text{refine}}$. Finally, the model $\Psi$ is optimized using $\mathcal{L}_{\text{LDL}}$, and the parameters of $\Psi$ are temporally ensembled to $\Psi_{\text{EMA}}$. This process is iterated until converge.
	
	With the proposed LDL, we train the same RRDB backbone~\cite{wang2018esrgan} and plot the MAD curves of intermediate GAN-SR outputs in Figure~\ref{curve} using dash lines. As can be seen, our LDL method has much better stability than ESRGAN in model learning, especially for type B and type C patches, resulting in much smaller MAD and MAD variations.
	
	\begin{table*}[t]\scriptsize
		\caption{Quantitative comparison between GAN-SR methods and the proposed LDL. Three groups of comparisons are made based on the employed backbone networks: SRResNet-like backbone for the first 3 columns, RRDB backbone for the middle 5, and SwinIR backbone for the last 2. The best results of each group are highlighted in \textbf{bold}. $\uparrow$ and $\downarrow$ mean that the larger or smaller score is better, respectively. }
		\vspace{-2em}
		\label{overall_table}
		\begin{center}
			\begin{tabular}{p{0.8cm}p{1.5cm}|p{1cm}<{\centering}p{1cm}<{\centering}p{1cm}<{\centering}|p{1cm}<{\centering}p{1cm}<{\centering}p{1cm}<{\centering}p{1cm}<{\centering}p{1cm}<{\centering}|p{1cm}<{\centering}p{1cm}<{\centering}}
				\toprule
				\textbf{}				Metrics&Benchmark&SFTGAN \cite{wang2018recovering}&SRGAN \cite{ledig2017photo}&SRResNet \cite{ledig2017photo}+LDL&ESRGAN \cite{wang2018esrgan}&USRGAN \cite{zhang2020deep}&SPSR \cite{ma2020structure}&RRDB \cite{wang2018esrgan}+LDL&RRDB\cite{wang2018esrgan}+LDL&SwinIR \cite{liang2021swinir}+$\mathcal{L}_{\text{GAN}}$ & SwinIR \cite{liang2021swinir}+LDL\\
				\midrule
				\multirow{2}{*}{Training Dataset}&&ImageNet + OST&DIV2K&DIV2K&DF2K + OST&DF2K&DIV2K&DIV2K&DF2K&DF2K&DF2K\\
				\midrule
				\multirow{6}*{LPIPS $\downarrow$}
				&Set5&0.0800&\textbf{0.0753}&0.0759&0.0758&0.0795&\textbf{0.0647}&0.0670&0.0691&0.0656&\textbf{0.0655}\\
				&Set14&0.1313&0.1327&\textbf{0.1303}&0.1241&0.1347&0.1207&0.1207&\textbf{0.1132}&0.1160&\textbf{0.1091}\\
				&Manga109&0.0716&0.0707&\textbf{0.0673}&0.0649&0.0630&0.0672&0.0553&\textbf{0.0544}&0.0542&\textbf{0.0469}\\
				&General100&0.0947&0.0964&\textbf{0.0898}&0.0879&0.0937&0.0862&\textbf{0.0790}&0.0796&0.0796&\textbf{0.0740}\\
				&Urban100&0.1343&0.1439&\textbf{0.1330}&0.1229&0.1330&0.1184&0.1096&\textbf{0.1084}&0.1077&\textbf{0.1021}\\
				&DIV2K100&0.1331&0.1257&\textbf{0.1172}&0.1154&0.1325&0.1099&0.1011&\textbf{0.0999}&0.1038&\textbf{0.0944}\\
				\midrule
				\multirow{6}*{DISTS $\downarrow$}
				&Set5&0.1085&\textbf{0.1003}&0.1010&0.0949&0.1045&0.0921&\textbf{0.0917}&0.0919&0.0930&\textbf{0.0899}\\
				&Set14&0.1133&0.1067&\textbf{0.1016}&0.0951&0.0997&0.0920&0.0935&\textbf{0.0866}&0.0930&\textbf{0.0869}\\
				&Manga109&0.0646&0.0557&\textbf{0.0523}&0.0471&0.0471&0.0463&0.0404&\textbf{0.0355}&0.0365&\textbf{0.0315}\\
				&General100&0.0992&0.0982&\textbf{0.0939}&0.0874&0.0931&0.0884&0.0827&\textbf{0.0801}&0.0835&\textbf{0.0794}\\
				&Urban100&0.1062&0.1081&\textbf{0.0989}&0.0880&0.0975&0.0849&0.0822&\textbf{0.0793}&0.0835&\textbf{0.0800}\\
				&DIV2K100&0.0736&0.0663&\textbf{0.0624}&0.0593&0.0645&0.0546&0.0528&\textbf{0.0526}&0.0531&\textbf{0.0507}\\
				\midrule
				\multirow{6}*{FID $\downarrow$}
				&Set5&39.261&31.507&\textbf{27.542}&27.215&37.006&30.904&25.288&\textbf{24.803}&35.401&\textbf{27.955}\\
				&Set14&60.493&63.945&\textbf{52.080}&54.933&55.635&53.867&49.577&\textbf{43.454}&48.910&\textbf{46.057}\\
				&Manga109&21.464&\textbf{11.948}&12.652&11.552&10.658&10.662&\textbf{9.855}&10.161&9.703&\textbf{8.680}\\
				&General100&36.845&33.868&\textbf{32.737}&29.843&32.959&30.159&27.506&\textbf{27.211}&27.557&\textbf{25.304}\\
				&Urban100&\textbf{21.370}&22.162&21.512&20.345&21.555&18.672&17.758&\textbf{16.351}&17.555&\textbf{16.282}\\
				&DIV2K100&18.183&\textbf{13.922}&14.823&13.557&14.031&13.754&12.145&\textbf{12.121}&12.736&\textbf{12.075}\\
				\midrule
				\multirow{6}*{PSNR $\uparrow$}
				&Set5&30.057&29.920&\textbf{30.527}&30.438&30.910&30.397&30.985&\textbf{31.033}&30.873&\textbf{31.028}\\
				&Set14&26.743&26.839&\textbf{27.278}&26.594&27.405&26.860&\textbf{27.491}&27.228&27.282&\textbf{27.526}\\
				&Manga109&28.167&28.110&\textbf{28.664}&28.413&28.753&28.561&29.407&\textbf{29.620}&29.345&\textbf{30.143}\\
				&General100&29.159&29.327&\textbf{29.775}&29.425&30.001&29.424&30.232&\textbf{30.289}&30.104&\textbf{30.441}\\
				&Urban100&24.338&24.410&\textbf{24.745}&24.365&24.891&24.804&\textbf{25.498}&25.459&25.736&\textbf{26.231}\\
				&DIV2K100&28.085&28.165&\textbf{28.602}&28.175&28.787&28.182&\textbf{28.951}&28.819&28.784&\textbf{29.117}\\
				\midrule
				\multirow{6}*{SSIM $\uparrow$}
				&Set5&0.8483&0.8478&\textbf{0.8570}&0.8523&0.8657&0.8443&\textbf{0.8626}&0.8611&\textbf{0.8655}&0.8611\\
				&Set14&0.7175&0.7252&\textbf{0.7366}&0.7144&0.7486&0.7254&\textbf{0.7476}&0.7358&0.7407&\textbf{0.7478}\\
				&Manga109&0.8562&0.8632&\textbf{0.8702}&0.8595&0.8717&0.8590&\textbf{0.8746}&0.8734&0.8796&\textbf{0.8880}\\
				&General100&0.8060&0.8074&\textbf{0.8164}&0.8095&0.8241&0.8091&0.8277&\textbf{0.8280}&0.8305&\textbf{0.8347}\\
				&Urban100&0.7235&0.7302&\textbf{0.7409}&0.7341&0.7503&0.7474&\textbf{0.7673}&0.7661&0.7786&\textbf{0.7918}\\
				&DIV2K100&0.7707&0.7745&\textbf{0.7855}&0.7759&0.7941&0.7720&\textbf{0.7951}&0.7897&0.7911&\textbf{0.8011}\\
				\bottomrule
			\end{tabular}
		\end{center}\vspace{-2.5em}
	\end{table*}
	
	\vspace{-0.3em}
	\section{Experimental results}
	\vspace{-0.3em}
	
	
	
	\subsection{Experiment setup}
	\vspace{-0.3em}
	
	\begin{figure*}[t]
		\centering
		\includegraphics[width=0.96\textwidth]{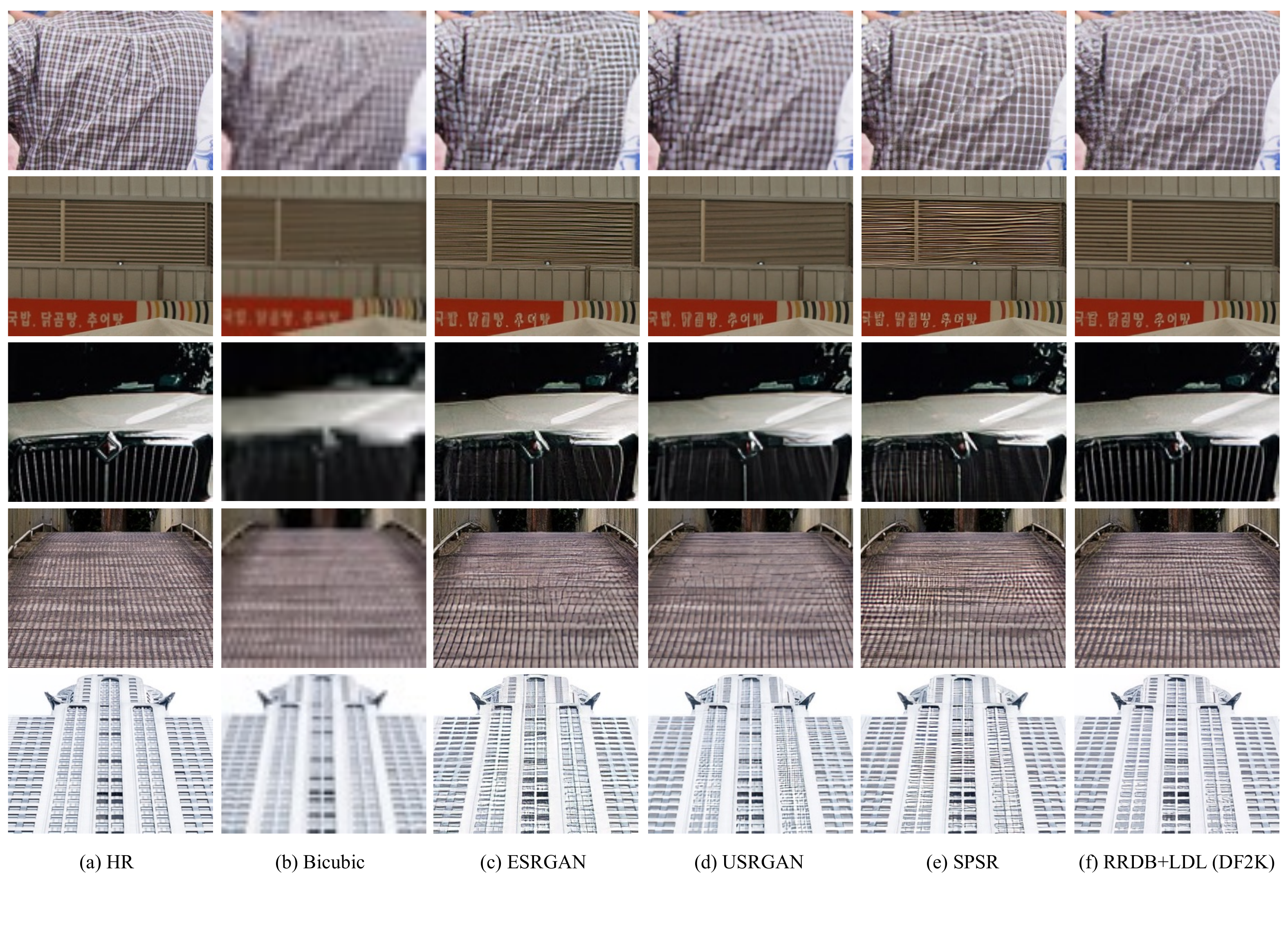}
		\vspace{-2.5em}
		\captionof{figure}{\label{Compare_all}Visual comparison (better zoom-in on screen) to state-of-the-art GAN-SR methods that use RRDB~\cite{wang2018esrgan} as backbone, including ESRGAN~\cite{wang2018esrgan}, USRGAN~\cite{zhang2020deep}, SPSR~\cite{ma2020structure} and our RRDB+LDL. As can be seen, our method has clear advantages in reconstructing realistic details and inhibiting artifacts. More visual comparisons can be found in the supplementary materials.}\vspace{-1em}
	\end{figure*}
	
	\textbf{Backbones and compared methods.} We validate the effectiveness of the proposed LDL method on top of three representative backbone networks, \ie, SRResNet~\cite{ledig2017photo}, RRDB~\cite{wang2018esrgan} and SwinIR~\cite{liang2021swinir}, resulting in SRResNet+LDL, RRDB+LDL and SwinIR+LDL. SRResNet is a light-weight network, and we compare SRResNet+LDL against SRGAN~\cite{ledig2017photo} and SFTGAN~\cite{wang2018recovering}, which have comparable number of parameters. RRDB is widely used in recent GAN-SR methods~\cite{wang2018esrgan, zhang2020deep, ma2020structure} for its competitive performance. We compare RRDB+LDL against ESRGAN~\cite{wang2018esrgan}, USRGAN~\cite{zhang2020deep} and SPSR~\cite{ma2020structure}, which all use RRDB as backbone. Very recently, SwinIR has reported excellent SISR performance by using the Swin Transformer architecture~\cite{liu2021swin}. We also train SwinIR with the $\mathcal{L}_{\text{LDL}}$ and $\mathcal{L}_{\text{GAN}}$  (SwinIR+$\mathcal{L}_{\text{GAN}}$) losses, respectively, and compare their performance. 
	We further validate LDL for real-world SISR by applying LDL to RealESRGAN~\cite{wang2021realesrgan}, and compare the obtained RealESRGAN+LDL model with both RealESRGAN and BSRGAN~\cite{zhang2021designing} models.
	
	\textbf{Training datasets and settings.} Following prior arts~\cite{ledig2017photo, wang2018esrgan, ma2020structure}, we conduct experiments with a scaling factor of $4\times$ on both synthetic (downsampled using MATLAB bicubic kernel) and real-world experiments. We also report $2\times$ GAN-SR results on synthetic data in the supplementary materials. We use the same data augmentation, discriminator and optimizer settings as in ESRGAN~\cite{wang2018esrgan}. We train our model on either DIV2K~\cite{agustsson2017ntire} ($800$ images) or DF2K ($3450$ images) dataset~\cite{lim2017enhanced, timofte2017ntire}, and the resolution of HR patches is $128\times 128$. We implement the experiments on $ 4 $ NVIDIA GTX 2080Ti GPUs with PyTorch and the batch size is $ 16 $ per GPU. We initialize the generator with a pretrained fidelity-oriented model, and calculate the perceptual loss as in~\cite{wang2021realesrgan} for both synthetic and real-world settings. The learning rate is $1e^{-4}$ and the number of training iteration is $300k$.
	
	\textbf{Evaluation benchmarks and metrics.} We employ $6$ benchmarks for evaluation, including Set5~\cite{bevilacqua2012low}, Set14~\cite{zeyde2010single}, Manga109~\cite{matsui2017sketch}, General100~\cite{dong2016accelerating}, Urban100~\cite{huang2015single} and DIV2K100~\cite{agustsson2017ntire}. We compare the GAN-SR results in terms of both perceptual quality and reconstruction accuracy. For the former, we employ LPIPS~\cite{zhang2018unreasonable}, DISTS~\cite{ding2020image} and FID~\cite{heusel2017gans} as metrics. LPIPS and DISTS have been validated effective on evaluating GAN-SR results~\cite{jinjin2020pipal}, and FID is widely used to evaluate the image perceptual quality in image generation tasks~\cite{karras2019style}. For the latter, we compute PSNR and SSIM indices on the Y channel in the YCbCr space. 
	
	\vspace{-0.2em}
	\subsection{Comparison with state-of-the-arts}
	\vspace{-0.2em}
	
	\textbf{Quantitative comparison.} Table~\ref{overall_table} compares quantitatively the state-of-the-art GAN-SR methods and our LDL. We can see that our proposed LDL scheme improves both the perceptual quality (LPIPS, DISTS, FID) and reconstruction accuracy (PSNR, SSIM) on most benchmarks under all the three backbones, \ie, SRResNet, RRDB and SwinIR. 
	
	Specifically, for the three light-weight models, SRResNet+LDL outperforms SFTGAN and SRGAN on most benchmarks in terms of those perceptual quality metrics LPIPS, DISTS and FID, and it outperforms SFTGAN and SRGAN on all benchmarks in terms reconstruction accuracy, \eg, PSNR +$0.3\sim0.5$dB and SSIM +$0.01$ over the second best method, respectively.

	\begin{figure*}[t]
		\centering
		\includegraphics[width=1\textwidth]{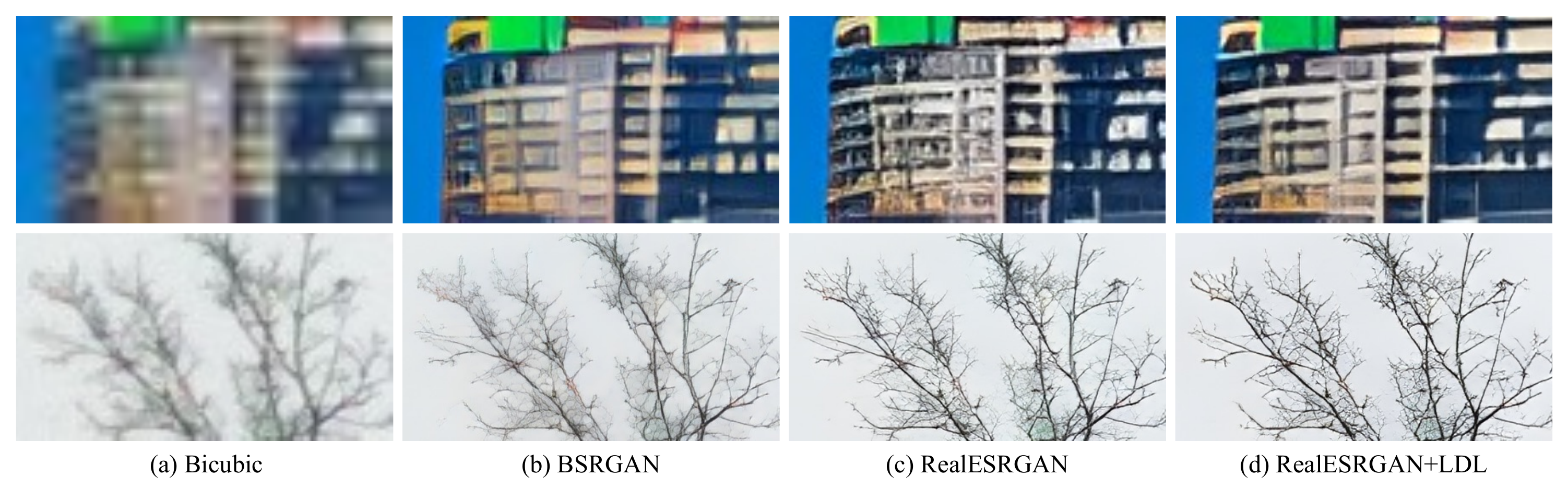}
		\vspace{-2em}
		\captionof{figure}{\label{Real}Visual comparison (better zoom-in on screen) to state-of-the-art real-world SISR methods, including BSRGAN~\cite{zhang2021designing} and RealESRGAN~\cite{wang2021realesrgan} . The training setting of RealESRGAN+LDL is the same as RealESRGAN except for the proposed $\mathcal{L}_{\text{LDL}}$ loss. More visual comparisons of different backbones can be found in the supplementary materials.}\vspace{-0.5em}
	\end{figure*}
	
	For the CNN based backbone RRDB, we train the GAN-SR models on DIV2K and DF2K, respectively, to be consistent with the employed competing models. We can see that among the three competing methods, SPSR performs the best in terms of perceptual quality metrics since it benefits from the additional network branch to restore the gradient map of images. By explicitly discriminating artifacts and regularizing the adversarial training, LDL achieves improvements against SPSR, \eg, LPIPS from $0.1099$ to $0.1011$ (about $8\%$) on DIV2K validation set. USRGAN achieves the best reconstruction accuracy among the three competing methods since it integrates learning-based and model-based strategies. Compared to the USRGAN, LDL not only achieves much better reconstruction accuracy on all benchmarks, but also improves the perceptual indexes. This validates that LDL can simultaneously inhibit the visual artifacts and generate more details with high-fidelity.
	
	For the transformer-based backbone SwinIR, we see that SwinIR+$\mathcal{L}_{\text{GAN}}$ outperforms the CNN based methods on most benchmarks in terms of both perceptual quality and reconstruction accuracy, demonstrating the potentials of transformer-based architecture for GAN-SR. As expected, SwinIR+LDL further improves SwinIR+$\mathcal{L}_{\text{GAN}}$ on most benchmarks, demonstrating the generalization capacity of LDL on different network architectures.
	
	\textbf{Qualitative comparison.} Figure~\ref{Compare_all} presents some visual comparisons among the GAN-SR methods using the RRDB backbone. Similar conclusions to the quantitative comparisons can be drawn. LDL generates much less visual artifacts compared to ESRGAN, USRGAN and SPSR, especially on regions with fine-scale aliasing structures. In addition, by regularizing the adversarial training process, LDL is able to reconstruct more details with high fidelity, such as the areas with regular patterns (\eg, the lines on windows and the grid on bridge). These improvements make LDL a practical GAN-SR solution for image quality enhancement.

	\begin{table}[t]\scriptsize
		\caption{\label{ablation}Ablation study on the different components of the proposed LDL method. Results are obtained by RRDB+LDL trained on DF2K and evaluated on DIV2K validation set. $\checkmark$ denotes that the corresponding operation is used.}
		\vspace{-2em}
		\label{SRCCs}
		\begin{center}
			\begin{tabular}{p{0.2cm}<{\centering}|p{0.7cm}<{\centering}p{0.7cm}<{\centering}p{0.7cm}<{\centering}p{0.7cm}<{\centering}|p{1cm}<{\centering}p{1cm}<{\centering}}
				\toprule
				\#&$ \bm{M} $&$\sigma\cdot\bm{M}$&$\bm{M}_{\text{refine}}$&$\Psi_{\text{EMA}}$&LPIPS&PSNR\\
				\midrule
				1&&&&&0.1154&28.175\\
				2&$\checkmark$&&&&0.1020&28.740\\
				3&&$\checkmark$&&&0.1006&28.678\\
				4&&&$\checkmark$&&0.1001&28.761\\
				5&&&$\checkmark$&$\checkmark$&0.0999&28.819\\
				\bottomrule
			\end{tabular}
		\end{center}\vspace{-2em}
	\end{table}

	\subsection{Applications to real-world SISR}
	
	To demonstrate the generalization capability of the proposed LDL, we also apply it to the real-world SISR task. Compared to SISR on synthetic LR images, SISR on real-world LR images faces unknown and much more complicated degradation~\cite{zhang2021designing}. We introduce the $\mathcal{L}_{\text{artif}}$ loss to the RealESRGAN method~\cite{wang2021realesrgan} and keep all other settings unchanged to train our RealESRGAN+LDL model. Since there is no ground-truth, we show qualitative comparisons with RealESRGAN and BSRGAN in Figure~\ref{Real}. As can be seen in the area of dense windows, RealESRGAN introduces unpleasant artifacts, while BSRGAN produces relatively smooth structures. In contrast, our LDL suppresses the generation of artifacts and encourages sharp details. In the area of twigs, the proposed LDL improves the generation of fine details, benefiting from the explicit and accurate discrimination between artifacts and realistic details.
	
	\subsection{Ablation study}
	
	We conduct ablation studies to investigate the roles of major components in our LDL method, including the primary artifact map $\bm{M}$, the globally scaled map $\sigma\cdot\bm{M}$ in Eq.~\eqref{var_global}, the refined map $\bm{M}_{\text{refine}}$ in Eq.~\eqref{refine} and the EMA model $\Psi_{\text{EMA}}$. Results are reported in Table~\ref{ablation}. \#1 gives the baseline performance when none of the above operations is used. By introducing $\bm{M}$ in \#2, we can observe a clear performance gain in both perceptual quality and reconstruction accuracy. This demonstrates the effectiveness of explicitly discriminating and penalizing the visual artifacts in GAN-SR. The usage of $\sigma\cdot\bm{M}$ in \#3 and $\bm{M}_{\text{refine}}$ in \#4 each further improves the performance. Finally, by using the stable EMA model $\Psi_{\text{EMA}}$ during testing in \#5, we achieve more performance gain as expected.
	
	\subsection{Limitations}
	
	Although the proposed LDL is effective in improving both the perceptual quality and reconstruction accuracy of SISR outputs, it still has some limitations in discriminating the visual artifacts in regions suffering from heavy aliasing. Take the last row of Figure~\ref{Compare_all} for example, there still remain some artifacts around the dense windows in our result. In this paper, we discussed how the artifacts are generated by GAN-SR methods and proposed a simple attempt to tackle this problem, while we believe there exist more effective designs for artifacts discrimination and details generation. 
	
	\section{Conclusion}
	
	In this paper, we analyzed how the visual artifacts were generated in the GAN-based SISR methods, and proposed a locally discriminative learning (LDL) strategy to address this issue. A framework to discriminate visual artifacts from realistic details during the GAN-SR model training process was carefully designed, and an artifact map was generated to explicitly penalize the artifacts without sacrificing the realistic details. The proposed LDL method can be easily plugged into different off-the-shelf GAN-SR models for both synthetic and real-world SISR tasks. Extensive experiments on the widely used datasets demonstrated that LDL outperforms the existing GAN-SR methods both quantitatively and qualitatively.
	
	\clearpage
	
	{\small
		\bibliographystyle{ieee_fullname}
		\bibliography{bib_LDL_CVPR2022}
	}
	
\end{document}